\documentclass[english,twocolumn,superscriptaddress]{revtex4-1}
\usepackage[T1]{fontenc}
\usepackage[latin9]{inputenc}
\usepackage{color}
\usepackage{amssymb}
\usepackage{graphicx}

\makeatletter
\@ifundefined{textcolor}{}
{%
 \definecolor{BLACK}{gray}{0}
 \definecolor{WHITE}{gray}{1}
 \definecolor{RED}{rgb}{1,0,0}
 \definecolor{GREEN}{rgb}{0,1,0}
 \definecolor{BLUE}{rgb}{0,0,1}
 \definecolor{CYAN}{cmyk}{1,0,0,0}
 \definecolor{MAGENTA}{cmyk}{0,1,0,0}
 \definecolor{YELLOW}{cmyk}{0,0,1,0}
 }

\makeatother

\usepackage{babel}
\begin{document}

\title{Two-setting multi-site Bell inequalities for loophole-free tests
with up to 50\% loss}

\author{M. D. Reid }

\affiliation{Centre for Quantum-Atom Optics, Swinburne University of Technology,
Melbourne, Australia}
\begin{abstract}
We consider Bell experiments with $N$ spatially separated qubits
where loss is present and restrict to two measurement settings per
site. We note the Mermin-Ardehali-Belinskii-Klyshko (MABK) Bell inequalities
do not present a tight bound for the predictions of local hidden variable
(LHV) theories. The Holder-type Bell inequality derived by Cavalcanti,
Foster, Reid and Drummond provides a tighter bound, for high losses.
We analyse the actual tight bound for the MABK inequalities, given
the measure $W=\prod_{k=1}^{N}\eta_{k}$ of overall detection efficiency,
where $\eta_{k}$ is the efficiency at the site $k$. Using these
inequalities, we confirm that the maximally entangled Greenberger-Horne-Zeilinger
state enables loophole-free falsification of LHV theories provided
$\prod_{k=1}^{N}\eta_{k}>2^{(2-N)}$, which implies a symmetric threshold
efficiency of $\eta\rightarrow50\%$ , as $N\rightarrow\infty$. Furthermore,
loophole-free violations remain possible, even when the efficiency
at some sites is reduced well below $0.5$, provided $N>3$.
\end{abstract}
\maketitle

\section{Introduction}

Bell showed the inconsistency of local realism with quantum mechanics,
by deriving a constraint on the correlations predicted by any local
hidden variable (LHV) theory \cite{bell}. For some quantum states,
these constraints, called Bell inequalities, are violated. Bell's
discovery of quantum nonlocality has inspired countless investigations
\cite{cs,clexp,aspect,weihs} and, through the close connection with
entanglement \cite{entperes}, underpins the field of quantum information.

A major challenge is to understand the interplay of Bell's nonlocality
with loss, which is defined by the ratio, $\eta$, of the number of
detected to emitted particles. Loss caused by detector inefficiencies
has resulted in the famous {}``detection loophole'' for testing
Bell's correlations in the laboratory \cite{loopholdet}. To date,
there has been no loophole-free violation of a Bell inequality for
spacelike separated measurement events. Furthermore, the sensitivity
of loophole-free Bell nonlocality to transmission losses is intimately
related to the security of quantum cryptography \cite{cry}. Motivated
by all this, there has been a considerable effort to work out the
smallest value of $\eta$ required for a loophole-free violation of
a Bell inequality.  

Bell's original gedanken experiment involved measurement of the spin
correlations of two maximally entangled and spatially separated spin-$1/2$
particles. His inequality, and the equally famous version derived
by Clauser-Horne-Shimony-Holt (CHSH), required only two measurement
settings, for each particle \cite{bell,bell2,chsh}. Despite the importance
of this inequality and its $N$-particle generalisations, the Mermin-Ardehali-Belinski-Klyshko
(MABK) inequalities \cite{mermin-1,ardehali,bk},  surprisingly little
is known about how to achieve a violation of them, for reduced efficiencies,
$\eta$. 

Where there are only two spatially separated particles ($N=2$), Garg
and Mermin put forward a modified CHSH inequality that could be violated
for $\eta>0.83$ \cite{gargmer}. Their inequality removed the necessity
of heralding the emission events, since the inequality did not specify
the total number of undetected particle pairs. Eberhard showed that
the Clauser-Horne (CH) inequality \cite{CH} would yield violations
for as low as $\eta>0.67$ using non-maximally entangled states, also
without the need for heralding \cite{ebernonmaxch}. Where measurements
are made on $N$ spin-$1/2$ systems, at each of $N$ sites, Larsson
and Semitecolos (LS) proved that for CH-type inequalities, $\eta>\frac{N}{2N-1}$
was sufficient, at least for some quantum state. They concluded {}``there
are $N$- site experiments for which the quantum mechanical predictions
violate local realism whenever $\eta>0.5$'' \cite{LS}. Despite
this knowledge, it remained unclear how to demonstrate this nonlocality,
as no specific inequality and state was proposed that would enable
realisation of Bell's nonlocality for efficiencies $\eta$ as low
as $0.5$ for each detector. 

 An explicit loophole-free demonstration of Bell nonlocality using
the MABK inequalities and the maximally entangled Greenberger-Horne-Zeilinger
(GHZ) states \cite{ghz} was shown possible, with heralding, for $\eta>2^{(1-N)/2N}$
\cite{braunmann}. However, the lowest threshold here requires $\eta>0.71$.
Cabello, Rodriguez and Villanueva (CRV) established that the LS limit
is achievable, for large $N$ and for the GHZ states, by proving that
$\eta>\frac{N}{2N-2}$ was necessary and sufficient for Bell nonlocality
in the case of $N$ odd. However, no inequality was proposed \cite{cabello-1}.
Firm proposals have been given however for efficiencies as low as
$\eta\rightarrow0.5$ at \emph{one} detector, but only for non-maximally
entangled states and provided an atom could be detected with 100\%
efficiency at a second site \cite{cablarasy,asymmbrunner}. 

In this paper, we contribute further to these results, by constructing
a tighter version of the MABK inequalities. Insight is gained from
the recent work of Cavalcanti et al and Acin et al \cite{cvbell2-1-1,vogelshc,acincavalcfrd,cvbell,prlholdcfrd,cfrdwig,cavalpranonlocality}
 who derived a {}``Holder'' Bell inequality that allows realisation
of the LS-CRV efficiency threshold of near $50\%$ for high $N$.
We will see that the tight MABK inequality is in fact a melding of
the new Holder inequality (which dominates at lower efficiencies)
and the old MABK inequality (which dominates at high efficiencies). 

In this way, we show it possible to violate a two-setting Bell inequality
loophole-free, using a maximally entangled GHZ state, whenever $\prod_{k=1}^{N}\eta_{k}>2^{(2-N)}$,
for $N\geq3$. Here, $\eta_{k}$ is the efficiency at site $k$.
For symmetric sites, the threshold efficiency reduces to $\eta\rightarrow0.5$
as $N\rightarrow\infty$, as given in Ref. \cite{cavalpranonlocality}.
In fact, as we confirm in this paper, $\eta\rightarrow0.5$ is the
best result possible, since we reason that $\eta>1/m$ is required
to demonstrate Bell's nonlocality using $m$-setting inequalities.
Furthermore, we establish that where $N>3$, the loophole-free violation
of the two-setting Bell inequality does not require $\eta_{k}>0.5$
for each site $k$, but can be achieved even if the efficiencies are
very low at some sites.\textbf{\emph{}}\textbf{ }

We conclude with a brief discussion, pointing out that three or more
sites are required if one is to obtain the violations of the two-setting
inequalities in the lossy scenarios. We then conjecture whether these
violations can signify a genuine multi-partite Bell nonlocality, in
the sense defined by Svetlichny and Collins et al \cite{svet}.

\section{Holder Bell inequalities\textbf{\emph{ }}}

Let us begin by presenting the Bell inequalities derived by Cavalcanti,
Foster, Reid and Drummond (CFRD) \cite{cvbell2-1-1}. We define a
set of space-like separated measurements $\hat{X_{\theta_{k}}^{k}}$
that can be performed on each of $k$ systems ($k=1,...,N$). We consider
only two settings, $\theta_{1}$ and $\theta_{2}$, at each site,
and so denote $\hat{X_{\theta_{1}}^{k}}=\hat{A_{k}}$ and $\hat{X_{\theta_{2}}^{k}}=\hat{B_{k}}$.
For any LHV theory, it is true that \cite{cvbell2-1-1,vogelshc,cvbell}
\begin{eqnarray}
\Bigl|\Bigl\langle\prod_{k=1}^{N}(A_{k}+iB_{k})\Bigr\rangle\Bigr|^{2} & \leq & \Bigl\langle\prod_{k=1}^{N}\{(A_{k})^{2}+(B_{k})^{2}\}\Bigr\rangle,\label{eq:bellcfrd}
\end{eqnarray}
where $A_{k}$, $B_{k}$ are the outcomes for the measurements $\hat{A_{k}}$
and $\hat{B_{k}}$ respectively. The left side of the inequality is
written in a compact form and involves moments of the Hermitian observables
$\hat{A_{k}}$ and $\hat{B_{k}}$ defined at each site. Violation
of this inequality will imply failure of LHV theories, and hence Bell's
nonlocality.

The inequality (\ref{eq:bellcfrd}) and its variants are closely associated
with the Holder inequalities used in mathematical analysis \cite{holder}.
For this reason, the inequalities based on (\ref{eq:bellcfrd}) will
be referred to throughout this paper as the {}``Holder Bell inequalities''.
The best known mathematical Holder inequality is the Cauchy Schwarz
inequality. The distinctive feature for our purposes is that the upper
bound given by the right-side of a Holder inequality is moment-dependent.
This will give an advantage for detecting Bell's nonlocality in lossy
scenarios.

The authors of Refs. \cite{acincavalcfrd,cavalpranonlocality} have
derived the application of the inequality (\ref{eq:bellcfrd}) to
the scenario of $N$ spin-$1/2$ systems, as in the original Bell
and GHZ gedanken experiments \cite{bell,ghz}. Here, one assigns $\hat{X_{\theta}^{k}}=\hat{\sigma_{k}}^{\theta}=\hat{\sigma_{k}}^{x}\cos\theta+\hat{\sigma_{k}}^{y}\sin\theta$,
where $\hat{\sigma_{k}}^{x/y}$ are the Pauli spin operators for site
$k$, and $\theta$ can be different for each site. Since the outcomes
of the measurement are always $+1$ or $-1$, the inequality (\ref{eq:bellcfrd})
will reduce to \cite{acincavalcfrd,cavalpranonlocality} 
\begin{eqnarray}
\Bigl|\Bigl\langle\prod_{k=1}^{N}(A_{k}+iB_{k})\Bigr\rangle\Bigr| & \leq & 2^{N/2}.\,\label{eq:mermin-4}
\end{eqnarray}
When the combination of moments given by the left side of the inequality
exceeds $2^{N/2}$, one can claim failure of LHV models. 

Supposing there is inefficient detection, we follow Bell's analysis
\cite{bell2,cs,gargmer} and note that for each emission event, the
{}``spin'' measurement made on each particle will have three possible
outcomes, depending on whether the spin is measured {}``up'', {}``down'',
or if there is {}``no detection''. The three outcomes are assigned
the numerical result $+1$, $-1$ and $0$, respectively. That is,
each $A_{k}$ and $B_{k}$ can now have values $\pm1$ or $0$. Then,
we note that the inequality deduced from (\ref{eq:bellcfrd}) changes.
While the inequality (\ref{eq:mermin-4}) is still valid, it is too
restrictive. The moments of the right-side are no longer necessarily
given by $2^{N/2}$ as in the perfect efficiency case of (\ref{eq:mermin-4}),
but can be measured, and compared with those of the left-side, to
give a more sensitive test for failure of LHV theories. 

In practice, assuming a detection efficiency $\eta_{k}$ for both
measurements ($A_{k}$ and $B_{k}$) at site $k$, the right-side
of the inequality (\ref{eq:mermin-4}) is predicted to be $2^{N/2}(\eta_{1}\eta_{2}...\eta_{N})^{1/2}$.
The inequality reduces to
\begin{eqnarray}
\Bigl|\Bigl\langle\prod_{k=1}^{N}(A_{k}+iB_{k})\Bigr\rangle\Bigr| & \leq & 2^{N/2}(\eta_{1}\eta_{2}...\eta_{N})^{1/2}\,.\label{eq:mermin-4-1}
\end{eqnarray}
This gives us the desired result, that the Bell inequality has a LHV
bound (given by the right-side) that reduces with efficiency $\eta_{k}$.

Next, we establish the connection with the well-known MABK Bell inequalities,
by noting there are two different forms of the Holder Bell inequalities
(\ref{eq:bellcfrd}) and (\ref{eq:mermin-4-1}). If $x^{2}+y^{2}=r^{2}$,
then it is always true that $x+y\leq\sqrt{2}r$, where $x,\, y,\, r$
are real numbers. Thus, on separating $\langle\prod_{k=1}^{N}(A_{k}+iB_{k})\rangle$
into real and imaginary parts (denoted by symbols $\Re e$ and $\Im m$),
the Holder Bell inequality (\ref{eq:bellcfrd}) implies the Bell inequality
\begin{eqnarray}
Ar_{N} & \equiv & s_{R}\Re e\Bigl\langle\prod_{k=1}^{N}(A_{k}+iB_{k})\Bigr\rangle+s_{I}\Im m\Bigl\langle\prod_{k=1}^{N}(A_{k}+iB_{k})\Bigr\rangle\nonumber \\
 &  & \leq\sqrt{2}\Bigl\langle\prod_{k=1}^{N}\{A_{k}^{2}+B_{k}^{2}\}\Bigr\rangle^{1/2}\label{eq:ardhold}
\end{eqnarray}
(where $s_{R},s_{I}=\pm1$). Using the reasoning explained above,
this inequality reduces to $ $ 
\begin{equation}
Ar_{N}\leq2^{(N+1)/2}(\eta_{1}\eta_{2}...\eta_{N})^{1/2}\label{eq:ardloss}
\end{equation}
for the lossy experiment, which gives a useful version of the inequality
(\ref{eq:mermin-4-1}). $ $The inequalities (\ref{eq:ardhold}) and
(\ref{eq:ardloss}) have the same left-side as the subset of MABK
inequalities called the Ardehali inequalities \cite{ardehali}, and
we have therefore denoted the left-side by the symbol $Ar_{N}$.

Also, following directly from (\ref{eq:bellcfrd}), because for any
complex number $z=\Re ez+i\Im mz$ it is true that $\Re ez,\Im mz\leq|z|$,
it follows that for any LHV model 
\begin{eqnarray}
M_{N} & \leq & \Bigl\langle\prod_{k=1}^{N}\{A_{k}^{2}+B_{k}^{2}\}\Bigr\rangle^{1/2}\label{eq:merhold}
\end{eqnarray}
where we can select $M_{N}$ to be either of $\Re e\langle\prod_{k=1}^{N}(A_{k}+iB_{k})\rangle$
or $ $$\Im m\langle\prod_{k=1}^{N}(A_{k}+iB_{k})\rangle$. In the
presence of losses, the inequality (\ref{eq:merhold}) becomes
\begin{equation}
M_{N}\leq2^{N/2}(\eta_{1}\eta_{2}...\eta_{N})^{1/2}\label{eq:merloss}
\end{equation}
which gives a second useful version of the inequality (\ref{eq:mermin-4-1}).
$ $ In this case, the inequalities have the same left-side as the
subset of MABK inequalities derived by Mermin \cite{mermin-1}, and
we have therefore denoted the left-side by the symbol $M_{N}$. 

The Bell inequalities (\ref{eq:ardhold}) and (\ref{eq:merhold})
were derived, from a different perspective, by Cavalcanti et al \cite{cavalpranonlocality}.
We will show below that while the two Holder Bell inequalities given
by (\ref{eq:ardhold}) and (\ref{eq:merhold}) have the same left-side
as the MABK Bell inequalities, the right-side is different.

\section{MABK Bell inequalities}

The left-side of the Holder Bell inequalities (\ref{eq:ardhold})
and (\ref{eq:merhold}) corresponds precisely to that used in the
well known Bell inequalities of MABK \cite{mermin-1,ardehali,bk}.
We now present the MABK Bell inequalities. In the MABK case, a different
bound is obtained for the LHV prediction. When $\eta_{k}=1$, this
bound is clearly tighter than that derived for the Holder inequalities. 

The MABK inequalities consist of two subsets, one for even $N$ and
one for odd $N$. The well-known {}``Ardehali'' MABK Bell inequality
applies only to even $N$, and is \cite{ardehali}
\begin{eqnarray}
Ar_{N} & \leq & 2^{N/2}.\label{eq:mabkard}
\end{eqnarray}
When $N=2$, the left side becomes
\begin{equation}
S=Ar_{2}=\langle A_{1}A_{2}\rangle-\langle B_{1}B_{2}\rangle+\langle A_{1}B_{2}\rangle+\langle B_{1}A_{2}\rangle\label{eq:chsh}
\end{equation}
 and Ardehali's inequality reduces to the well-known CHSH inequality,
$S\leq2$. For the case of $N$ odd, only, Mermin proved the Bell
inequality \cite{mermin-1}
\begin{eqnarray}
M_{N} & \leq & 2^{(N-1)/2}\,.\label{eq:mermin-3}
\end{eqnarray}
Combined, the two inequalities (\ref{eq:mabkard}) and (\ref{eq:mermin-3})
give a LHV prediction for arbitrary $N$, and are commonly termed
the {}``MABK inequalities'' \cite{mermin-1,ardehali,bk}.

The Ardehali and Mermin Bell inequalities are also valid for the lossy
scenario, where {}``no detection'' outcomes are assigned the outcome
{}``$0$'' \cite{bell2,braunmann}. However, we can see immediately
on comparison with the Holder Bell inequalities (\ref{eq:ardloss})
and (\ref{eq:merloss}) that, for the lossy experiment, the MABK inequalities
can no longer be tight. A similar result is known for these inequalities
even in the context of pure states: MABK inequalities do not detect
the Bell nonlocality that has been shown to exist for nonmaximally
entangled generalised GHZ states \cite{mabknottight}.

\section{Quantum predictions }

Now, we examine the predictions given by the maximally entangled GHZ
state $\frac{1}{\sqrt{2}}\{|\uparrow\rangle^{\otimes N}-|\downarrow\rangle^{\otimes N}\}$,
where $|\uparrow\rangle_{k}$, $|\downarrow\rangle_{k}$ are the eigenstates
of the Pauli spin observable $\hat{\sigma}_{k}^{z}$,\textcolor{red}{{}
}\textcolor{black}{and $|\uparrow\rangle^{\otimes N}\equiv\prod_{k=1}^{N}|\uparrow\rangle_{k}$
\cite{wernerwolfmaxghz-2}.} Mermin showed that if we choose measurement
settings to correspond to either $ $$\hat{\sigma}_{k}^{x}$ or $\hat{\sigma}_{k}^{y}$
for each $j=1,...,N$, then for optimally chosen settings, the quantum
prediction is $M_{N}=2^{N-1}$, for $N$ odd \cite{mermin-1}. For
the Ardehali's inequalities, the optimal measurement choice involves
a $ $$\hat{\sigma}_{k}^{x}$ or $\hat{\sigma}_{k}^{y}$ setting for
$N-1$ sites, with a rotated setting for the $N$th site \cite{ardehali}.
Then, the optimal quantum prediction is $Ar_{N}=2^{N-1/2}$, for $N$
even. Assuming symmetric detector efficiencies $\eta_{k}=\eta$, the
optimal quantum prediction in the lossy case will be $S=\eta^{2}2\sqrt{2}$,
$M_{N}=\eta^{N}2^{N-1}$ (for odd $N$), and $Ar_{N}=\eta^{N}2^{N-1/2}$
(for $N$even). Using the MABK Bell inequalities directly, this gives
the efficiency threshold $\eta>2^{(1-N)/2N}$ for all $N$, which
reduces to a lowest value of $\eta\rightarrow0.71$ as $N\rightarrow\infty$,
as shown by Braunstein and Mann \cite{braunmann}. 

As pointed out in Ref. \cite{cavalpranonlocality}, the Holder Bell
inequalities (\ref{eq:ardhold}) and (\ref{eq:merhold}) will give
a lower efficiency threshold in the symmetric case, for all $N>3$.
If we consider odd $N$, then we use the inequality (\ref{eq:merhold}),
for which the right-side is predicted to be $2^{N/2}\eta^{N/2}$ (corresponding
to Eq. (\ref{eq:merloss})). For even $N$, we use the inequality
(\ref{eq:ardhold}) for which the right-side is $2^{(N+1)/2}\eta^{N/2}$
(corresponding to Eq. (\ref{eq:ardloss})). The associated threshold
efficiency for violation of the Holder-Bell inequalities is given
by $\eta>2^{(-1+2/N)}$ for all $N\geq3$ \cite{cavalpranonlocality}.
This threshold reduces to $0.79$ for $N=3$, and approaches $0.5$
as $N\rightarrow\infty$.

\section{LHV theory predictions}

Having confirmed that neither the MABK nor the Holder Bell inequalities
can provide the tight LHV bound in the presence of loss (poor detection
efficiencies), our objective is to gain insight into the \emph{actual}
LHV predictions, and to then determine if lower efficiency thresholds
are possible for a given $N$. 

For all LHV theories, it is true that \cite{bell} 
\begin{equation}
E(X_{\theta_{1}}^{1},...,X_{\theta_{N}}^{N})=\int_{\lambda}p(\lambda)E_{\lambda}(X_{\theta_{1}}^{1})...E_{\lambda}(X_{\theta_{n}}^{N})d\lambda\,\label{eq:bell-1}
\end{equation}
where $E(X_{\theta_{1}}^{1},...,X_{\theta_{N}}^{N})$ is the expectation
value for the product of outcomes $X_{\theta_{k}}^{k}$ of simultaneous
measurements $\hat{X_{\theta_{k}}^{k}}$ ($k=1,...,N$) performed
on the $N$ spatially separated systems. Here, $\{\lambda\}$ symbolises
the set of local hidden variables of the LHV theory. Thus, $E_{\lambda}(X_{\theta_{k}}^{k})$
is the expectation value of $X_{\theta_{k}}^{k}$ given the hidden
variable specification $\{\lambda\}$, and $p(\lambda)$ is the underlying
probability distribution for $\{\lambda\}$. 

We consider the LHV prediction for the terms $S$, $M_{N}$, $Ar_{N}$
of the CHSH, Mermin and Ardehali inequalities. We also introduce $W_{N}$
as a measure of overall efficiency. Specifically: 
\begin{eqnarray}
W_{N} & = & \frac{1}{2^{N}}\langle\prod_{k=1}^{N}\{|A_{k}|+|B_{k}|\}\rangle\label{eq:wset}
\end{eqnarray}
Where the outcomes of $A_{k}$ and $B_{k}$ are given by $+1$, $-1$
or $0$, it is clear that $W_{N}=\frac{1}{2^{N}}R$, where $R$ is
the right side of the Holder inequalities (\ref{eq:ardhold}) and
(\ref{eq:merhold}). In fact, there are $2^{N}$ relevant efficiencies
$\eta$, one for each measurement setting ($A_{k}$ or $B_{k}$) at
each site $k$. Where the efficiencies are equal for the two settings,
and given at site $k$ by $\eta_{k}$, the quantum prediction for
(\ref{eq:wset}) is $W_{N}=\prod_{k=1}^{N}\eta_{k}$. A complication
is that to measure the actual values of $S$, $M_{N}$, $Ar_{N}$
and $W_{N}$, it is necessary to establish all emission events, using
an {}``event ready'' or {}``heralding'' apparatus \cite{bell2,cs}.
This is a significant but not insurmountable challenge \cite{herald}.\textcolor{red}{{} }

It is possible to show that for any LHV model  \cite{bell2,cs}
\begin{equation}
\langle S\rangle=\int_{\lambda}p(\lambda)S_{\lambda}d\lambda,\,\,\,\,\,\,\,\,\langle W_{N}\rangle=\int_{\lambda}p(\lambda)W_{N,\lambda}d\lambda.\label{eq:lhvset}
\end{equation}
Similar expressions exist for $\langle M_{N}\rangle$ and $\langle Ar_{N}\rangle$.
Here, $S_{\lambda}=\langle A_{1}\rangle_{\lambda}\langle A_{2}\rangle_{\lambda}-\langle A_{1}\rangle_{\lambda}\langle B_{2}\rangle_{\lambda}+\langle B_{1}\rangle_{\lambda}\langle A_{2}\rangle_{\lambda}+\langle B_{1}\rangle_{\lambda}\langle B_{2}\rangle_{\lambda}$
and $W_{2,\lambda}=\langle|A_{1}|\rangle_{\lambda}\langle|A_{2}|\rangle_{\lambda}+\langle|A_{1}|\rangle_{\lambda}\langle|B_{2}|\rangle_{\lambda}+\langle|B_{1}|\rangle_{\lambda}\langle|A_{2}|\rangle_{\lambda}+\langle|B_{1}|\rangle_{\lambda}\langle|B_{2}|\rangle_{\lambda}$,
where $\langle A_{k}\rangle_{\lambda}$, $\langle B_{k}\rangle_{\lambda}$
are the expectation values for $A_{k}$, $B_{k}$, given the hidden
variable specification $\{\lambda\}$. Similar expansions can be given
for $W_{N,\lambda}$, $M_{N,\lambda}$ and $Ar_{N,\lambda}$. We find
that for any LHV theory, constraints exist for the possible values
of $\langle S\rangle$, $\langle M_{N}\rangle$ and $\langle Ar_{N}\rangle$,
given the value of $\langle W_{N}\rangle$. In other words, for any
given experimentally measured value of $W_{N}$, there will be a constraint
on the LHV predictions for $\langle S\rangle$, $\langle M_{N}\rangle$
and $\langle Ar_{N}\rangle$.

We determine these constraints as follows. The outcome $A_{k}$ is
constrained to be one of $\pm1$ or $0$. Thus, in the LHV model,
it must be true that $-1\leq\langle A_{k}\rangle_{\lambda}\leq1$.
The LHV model will specify probabilities for the $+1$ and $-1$ outcome
for $A_{k}$, for a given hidden variable specification $\{\lambda\}$.
We denote these probabilities by $ $$P_{\lambda}^{A}(+)$ and $P_{\lambda}^{A}(-)$.
Then we see that $\langle A_{k}\rangle_{\lambda}=P_{\lambda}^{A}(+)-P_{\lambda}^{A}(-)$
whereas $\langle|A_{k}|\rangle_{\lambda}=P_{\lambda}^{A}(+)+P_{\lambda}^{A}(-)$.
We note $P_{\lambda}^{A}(+)+P_{\lambda}^{A}(-)$ is in fact the efficiency
value predicted for the measurement setting $A_{k}$, given the hidden
variable specification $\{\lambda\}$, and we introduce the notation
$\langle\eta_{k}^{A}\rangle_{\lambda}\equiv\langle|A_{k}|\rangle_{\lambda}=P_{\lambda}^{A}(+)+P_{\lambda}^{A}(-)$.
Thus, for a given $\langle|A_{k}|\rangle_{\lambda}$, it follows that
\begin{equation}
-\langle\eta_{k}^{A}\rangle_{\lambda}\leq\langle A_{k}\rangle\leq\langle\eta_{k}^{A}\rangle_{\lambda}\label{eq:rect1}
\end{equation}
 and similarly
\begin{equation}
-\langle\eta_{k}^{B}\rangle_{\lambda}\leq\langle B_{k}\rangle\leq\langle\eta_{k}^{B}\rangle_{\lambda}.\label{eq:rect2}
\end{equation}
We evaluate for each possible $\langle\eta_{k}^{A}\rangle_{\lambda}$,
$\langle\eta_{k}^{B}\rangle_{\lambda}$, the possible values of $S_{\lambda}$
and $W_{\lambda}$, which is a simple numeric exercise. For a given
$W_{\lambda}$, the possible values of $S_{\lambda}$ can be displayed
as a scattering of points on a diagram. We can then sample again over
all possible distributions $p(\lambda)$ to evaluate the consistent
predictions for both $W_{2}$ and $S$, for any possible LHV theory
distribution. The same procedure is performed, to evaluate the possible
$M_{N}$ and $A_{N}$, for a given $W_{N}$. 

In fact, the full sampling is a tedious task. For our purposes, because
we have two analytical bounds on the LHV predictions, we sample LHV
predictions only to verify the bounds, and to establish the degree
of tightness of them. Our sampling involves evaluating the possible
$S$, $Ar_{N}$ or $M_{N}$, and $W_{N}$ when hidden variables assume
the extreme values of $\pm1$, or the value $0$, which in the absence
of loss would amount to assuming a deterministic LHV theory \cite{fine}.
This does not cover all stochastic LHV theories, but we will see is
enough to establish the validity and degree of tightness of the analytical
MABK and Holder limits, for a given $W_{N}$.

\section{Tightness of the inequalities}

Before analysing the results, we give a geometrical interpretation
of the degree of tightness of the Holder and MABK Bell inequalities.
The derivation of the MABK inequalities utilises that the local hidden
variable expectation values $\langle A_{k}\rangle_{\lambda}$, $\langle B_{k}\rangle_{\lambda}$
are each constrained to the domain $[-1,1]$. The MABK Bell inequality
is thus defined by the polytope formed from the $2$- dimensional
polytope, that is a square $\mathcal{S}$ centred at the origin, with
sides of length $2$ \cite{wernerwolfmaxghz-2}. 

The Holder inequality (\ref{eq:bellcfrd}) on the other hand is derived
using that for the local hidden variables, it is always true that
$\langle A_{k}\rangle_{\lambda}^{2}\leq\langle A_{k}^{2}\rangle$,
$\langle B_{k}\rangle_{\lambda}^{2}\leq\langle B_{k}^{2}\rangle$,
and hence that
\begin{equation}
\langle A_{k}\rangle_{\lambda}^{2}+\langle B_{k}\rangle_{\lambda}^{2}\leq\langle A_{k}^{2}\rangle_{\lambda}+\langle B_{k}^{2}\rangle_{\lambda}.\label{eq:ineqtightcircle}
\end{equation}
These constraints follow from the LHV assumption of a non-negative
variance for hidden variable distributions \cite{cvbell2-1-1,cvbell,cavalpranonlocality,vogelshc}.
For the case of perfect efficiency (corresponding to $W_{N}=1$),
this latter inequality reduces to $\langle A_{k}\rangle_{\lambda}^{2}+\langle B_{k}\rangle_{\lambda}^{2}\leq2$
(because the outcomes $A_{k}$, $B_{k}$ are always $\pm1$). In the
Holder derivation therefore, the values for the local hidden variables
are assumed constrained on or within a circle $\mathcal{C}$ centred
at the origin of radius $\sqrt{2}$, which \emph{encloses} the MABK
square $\mathcal{S}$. Clearly, this Holder constraint is not as tight
as the MABK one, and indeed the Holder Bell inequalities are not as
tight in this perfect efficiency limit.

Where the quantity $W_{N}$ reduces below $1$, however, the right-side
of (\ref{eq:ineqtightcircle}) is reduced. In fact we have seen from
the analysis given in the previous Section that for the LHV variables,
$\langle A_{k}^{2}\rangle_{\lambda}\leq\langle|A_{k}|\rangle_{\lambda}=\langle\eta_{k}^{A}\rangle_{\lambda}$.
The Holder derivation assumes $\langle A_{k}\rangle_{\lambda}^{2}\leq\langle A_{k}^{2}\rangle$,
and hence $\langle A_{k}\rangle_{\lambda}^{2}\leq\langle\eta_{k}^{A}\rangle_{\lambda}$,
and, similarly, that $\langle B_{k}\rangle_{\lambda}^{2}\leq\langle\eta_{k}^{B}\rangle_{\lambda}$.
These constraints can be written $ $$\frac{}{}$$\frac{\langle A_{k}\rangle_{\lambda}^{2}}{\langle\eta_{k}^{A}\rangle_{\lambda}}+\frac{\langle B_{k}\rangle_{\lambda}^{2}}{\langle\eta_{k}^{B}\rangle_{\lambda}}\leq2$,
which is the geometric constraint that the hidden variable expectation
values be on or within the ellipse $\mathcal{E}$, centred at the
origin and with minor and major radii given by $\sqrt{2}\sqrt{\langle\eta_{k}^{A}\rangle_{\lambda}}$
and $\sqrt{2}\sqrt{\langle\eta_{k}^{B}\rangle_{\lambda}}$. We see
that this constraint can become less restrictive than the MABK square
$S$, the consequence being that the Holder inequalities can become
tighter than the MABK ones, for lower efficiencies. We note the tight
LHV bound for the hidden variables is in fact given by the rectangle,
defined by (\ref{eq:rect1}-\ref{eq:rect2}), which is enclosed by
the ellipse $\mathcal{E}$. As a result, we cannot prove that the
Holder inequalities are tight.

\section{Results}

The two Bell inequalities, the Holder and MABK,  apply to all LHV
theories, and hence constrain all LHV predictions. There are valid,
for any given measurement of $W_{N}$, without any additional assumptions.
The MABK inequalities are categorised into two subsets, one for even
$N$, and one for odd $N$, and we will do the same for the Holder
inequalities. Using the definition of $W_{N}$ given by (\ref{eq:wset}),
we can rewrite the Holder inequalities Eqs (\ref{eq:ardhold}), (\ref{eq:merhold}).
The two sets of inequalities are: $Ar_{N}\leq2^{(N+1)/2}\sqrt{W_{N}}$,
$M_{N}\leq2^{N/2}\sqrt{W_{N}}$ for the Holder case; and $Ar_{N}\leq2^{N/2}$
and $M_{N}\leq2^{(N-1)/2}$ for the MABK case (for even and odd $N$
respectively, in each case). 

The Holder and MABK upper bounds coincide when $W_{N}=0.5$. For $W_{N}<0.5$,
the Holder bound is tighter for establishing violations of LHV theories.
For all $W_{N}\geq0.5$, the MABK bound is the tighter bound. Results
of possible LHV predictions are plotted in Figures 1-4, as a scattering
of points in the graphs of $S$, $M_{N}$ or $Ar_{N}$ versus $W_{N}$,
or efficiency $\eta$. The Holder and MABK bounds contain below them
all the LHV predictions. 

We consider the experiment where the correlations are generated by
a maximally entangled GHZ state, and efficiencies at each site are
$\eta_{k}$. The quantum predictions are $W_{N}=\prod_{k=1}^{N}\eta_{k}$
and $S=2\sqrt{2}W_{2}$, $M_{N}=2^{N-1}W_{N}$, $Ar_{N}=2^{N-1/2}W_{N}$.
When $N=2$, this quantum prediction does not cross the Holder bound
\cite{acincavalcfrd}, as seen from Fig. 1, and a violation of the
CHSH Bell inequality requires $\eta_{1}\eta_{2}>1/\sqrt{2}$. 

\begin{figure}[h]
\begin{centering}
\includegraphics[width=1\columnwidth]{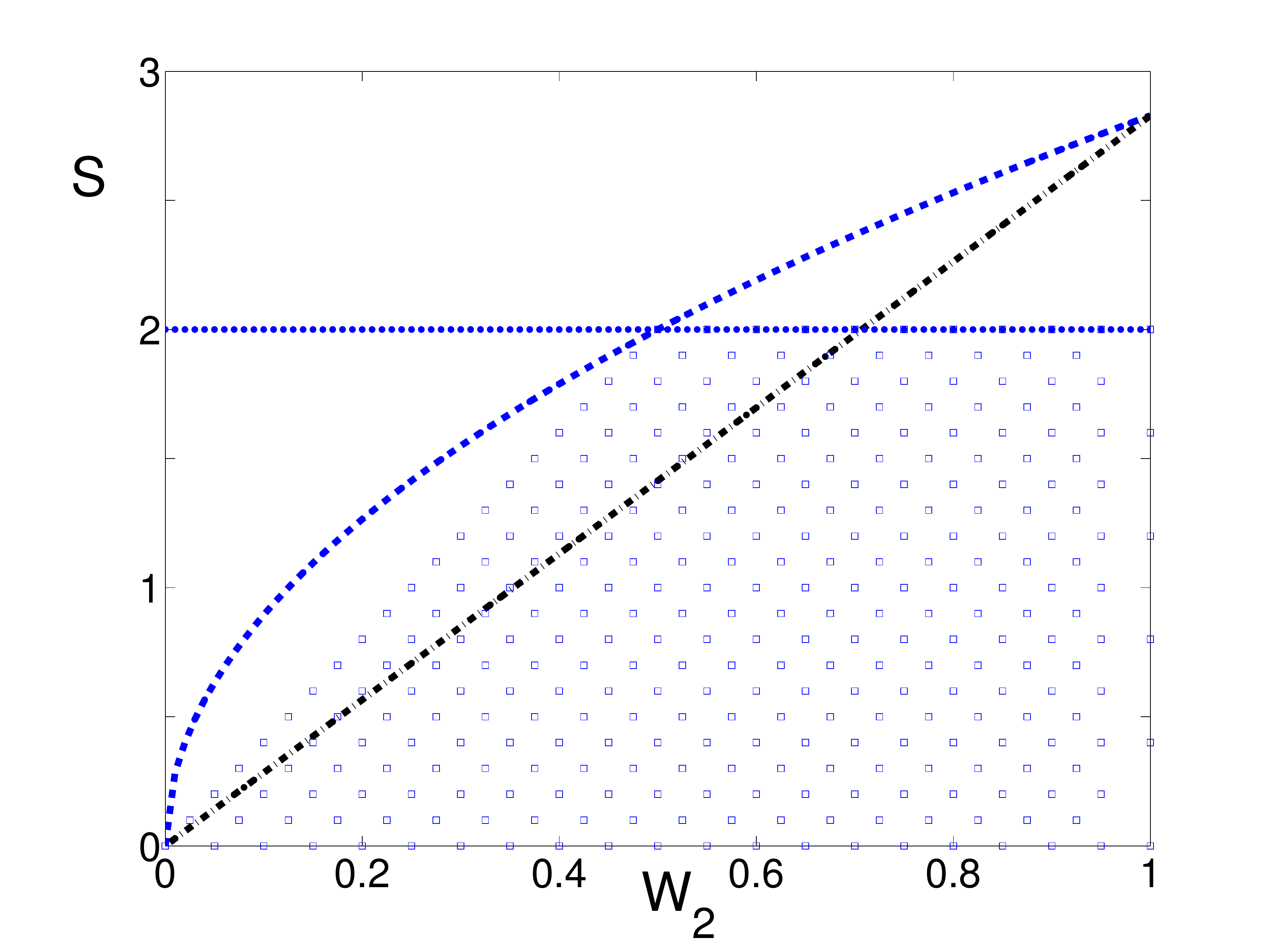} 
\par\end{centering}

\caption{The Local Hidden Variable (LHV) and quantum predictions for $S$,
for a given value of $W_{2}$ which is a measure of efficiency. The
blue dashed curve is the Holder inequality bound, which provides an
upper limit to the LHV prediction for $S$. The blue dotted line at
$S=2$ is the CHSH inequality bound, which also provides an upper
limit to the LHV prediction for $S$. The blue squares give a scattering
of actual predictions for LHV theories. The observation of $S$ greater
than either of the Holder or CHSH curves confirms the failure of all
LHV theories. The dashed black solid line gives the quantum prediction,
for the special case of a maximally entangled Bell state.}
\end{figure}

\begin{figure}[h]
\begin{centering}
 \includegraphics[width=1\columnwidth]{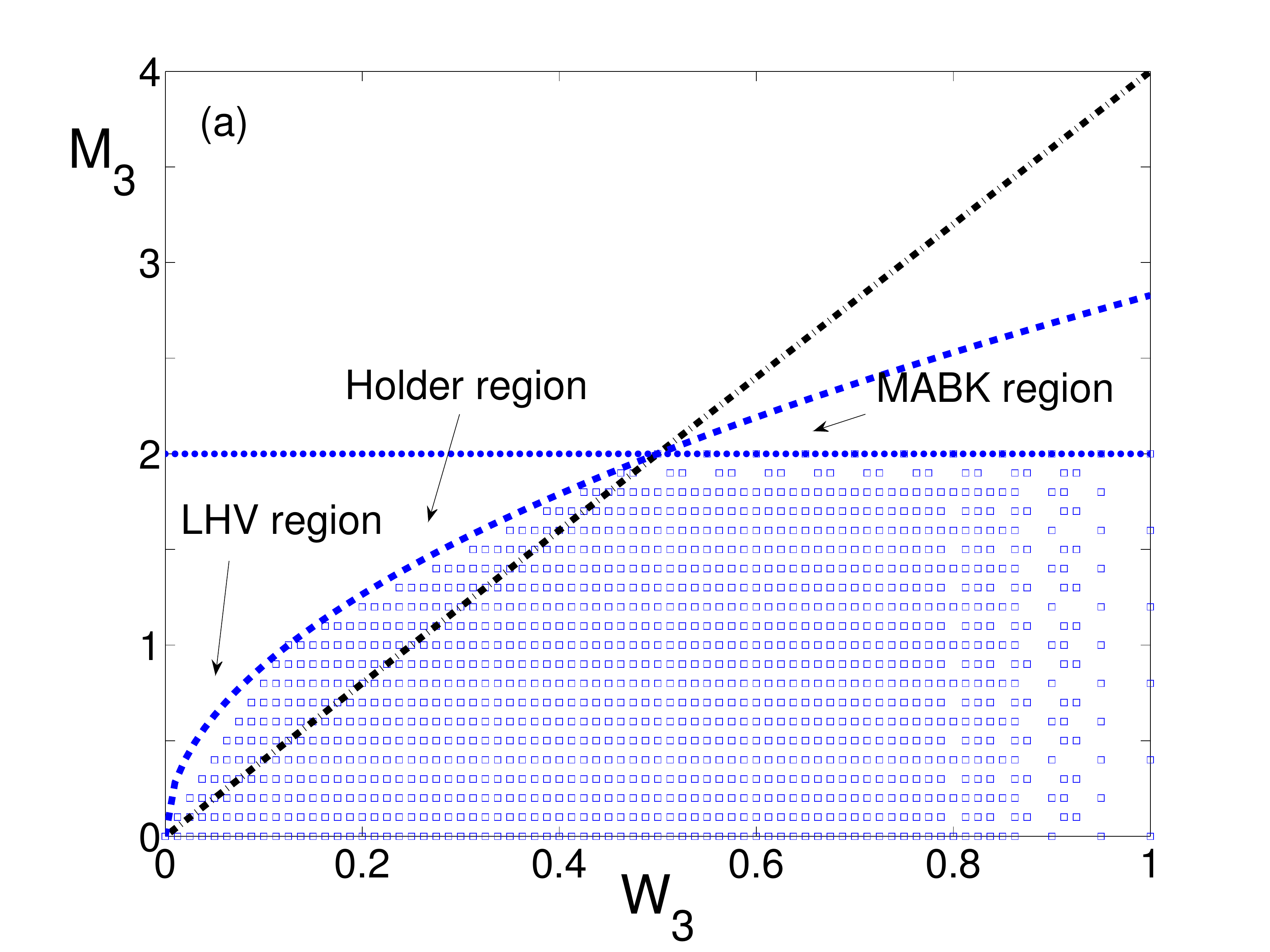} \includegraphics[width=1\columnwidth]{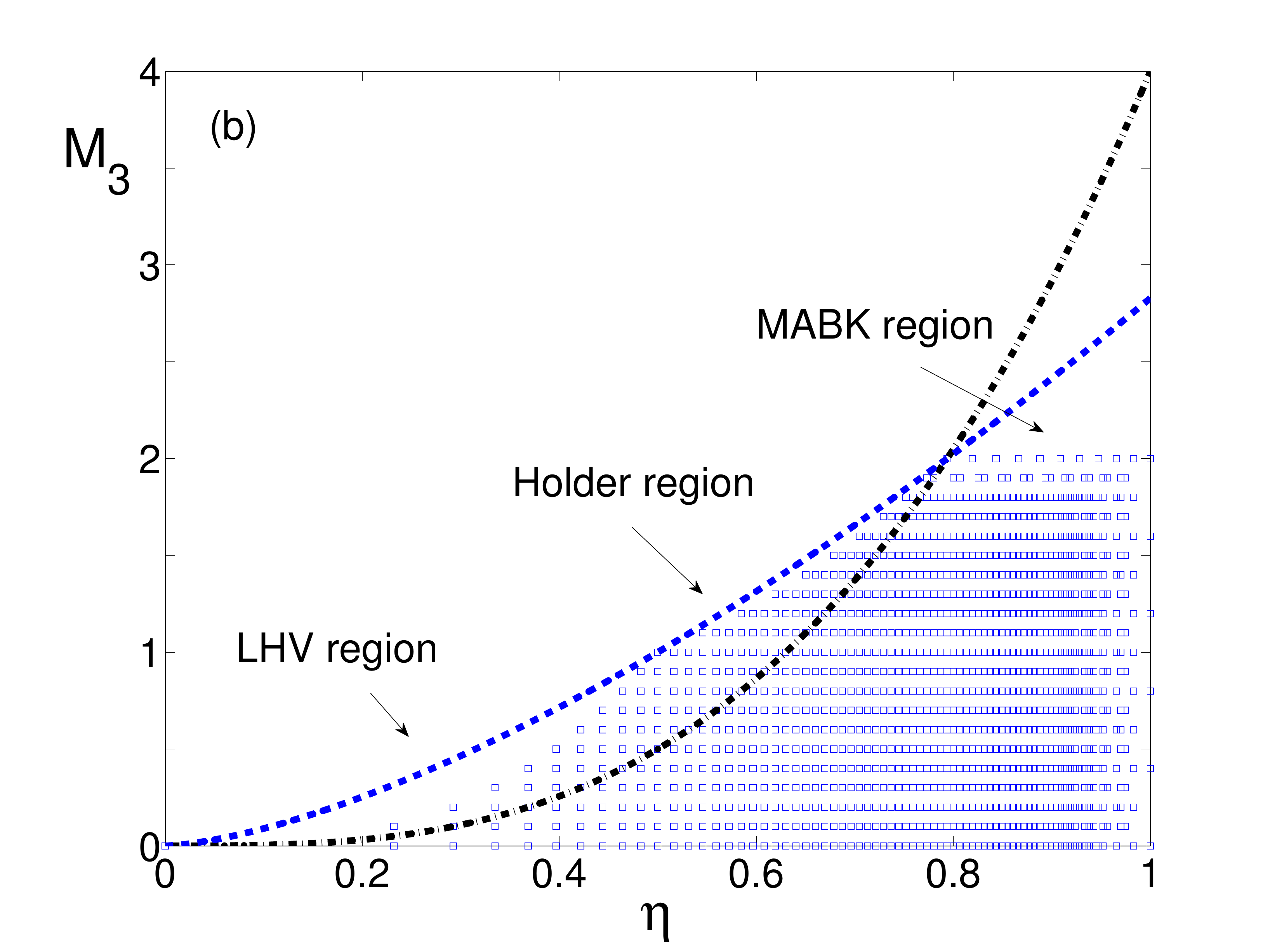} 
\par\end{centering}

\caption{The Local Hidden Variable (LHV) and quantum predictions for $M_{3}$,
versus efficiency. The top Figure plots $M_{3}$, for a specified
value of $W_{N}$. The blue dashed curve is the Holder inequality
bound, which gives an upper limit to the LHV predictions. The blue
dotted line is the upper LHV bound given by the MABK inequality. The
blue squares are a scattering of actual LHV predictions. The black
line is the quantum prediction, for the maximally entangled GHZ state.
The lower figure plots the predictions versus the actual efficiency
at each site, for the case of symmetric detection efficiencies where
$\eta_{k}=\eta$, $\forall k$.}
\end{figure}
 
\begin{figure}[h]
\begin{centering}
  \includegraphics[width=1\columnwidth]{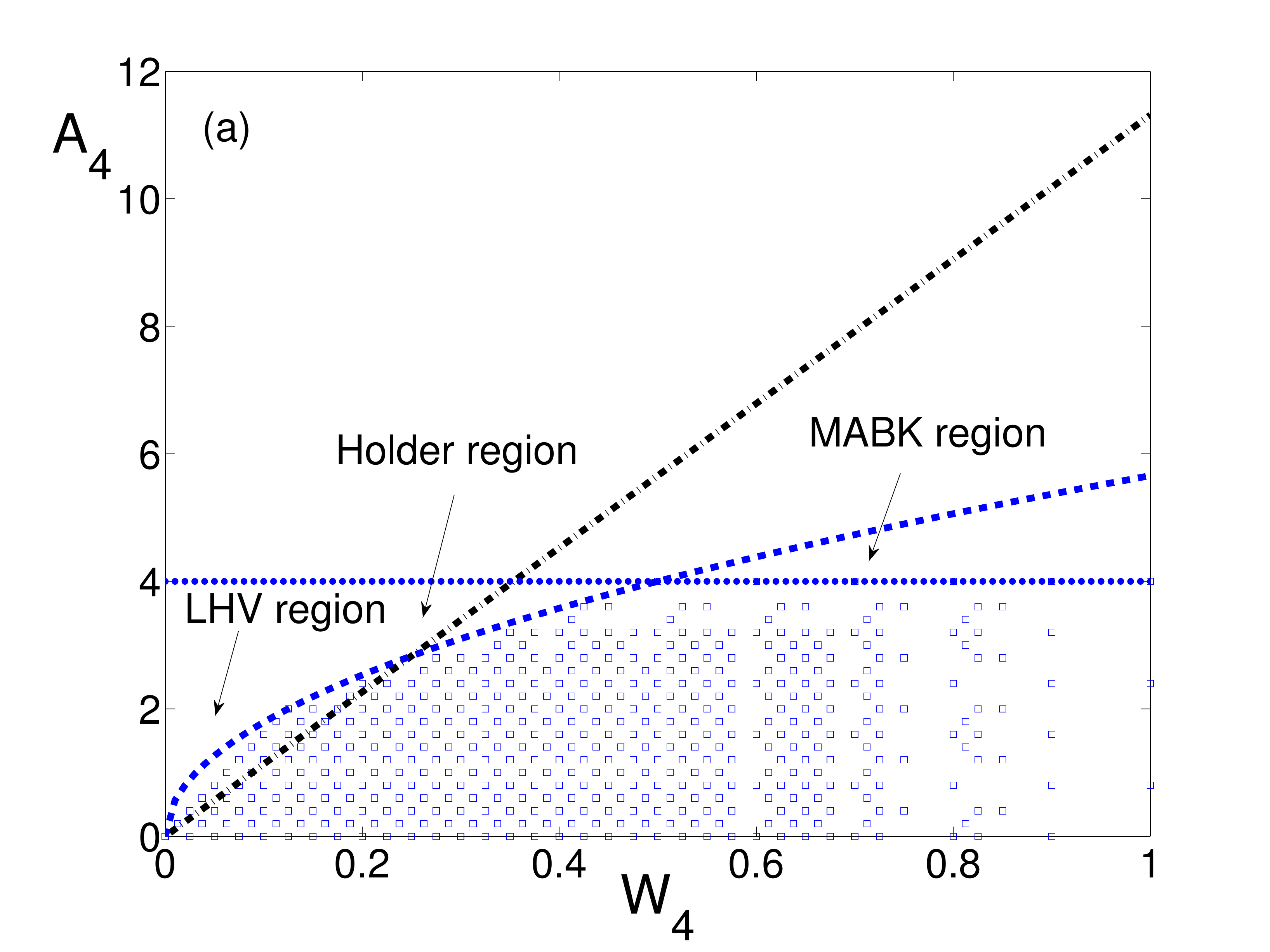} 
\par\end{centering}

\begin{centering}
\includegraphics[width=1\columnwidth]{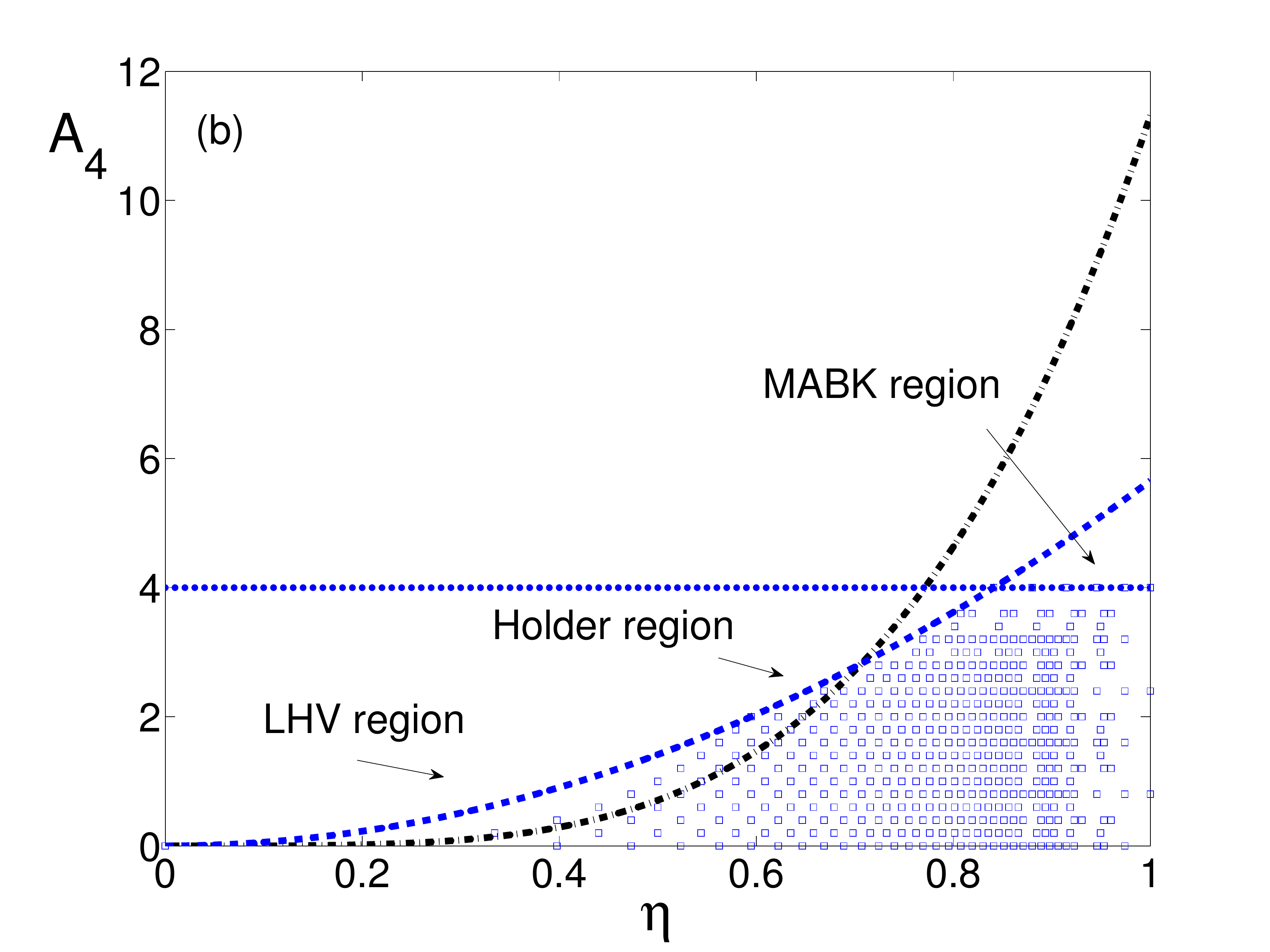}
\par\end{centering}

\caption{The LHV and quantum predictions for $Ar_{4}$ versus efficiency. The
curves are as defined in Figures 1 and 2. The top figure gives the
predictions for $Ar_{4}$, given a value of $W_{N}$. The lower figure
gives the predictions for the special symmetric case, where $\eta_{k}=\eta$,
$\forall k$. The blue curves are the Holder and MABK inequality bounds
to the LHV predictions. The black line is the quantum prediction for
a GHZ state. }
\end{figure}
\begin{figure}[h]
\begin{centering}
 \includegraphics[width=1\columnwidth]{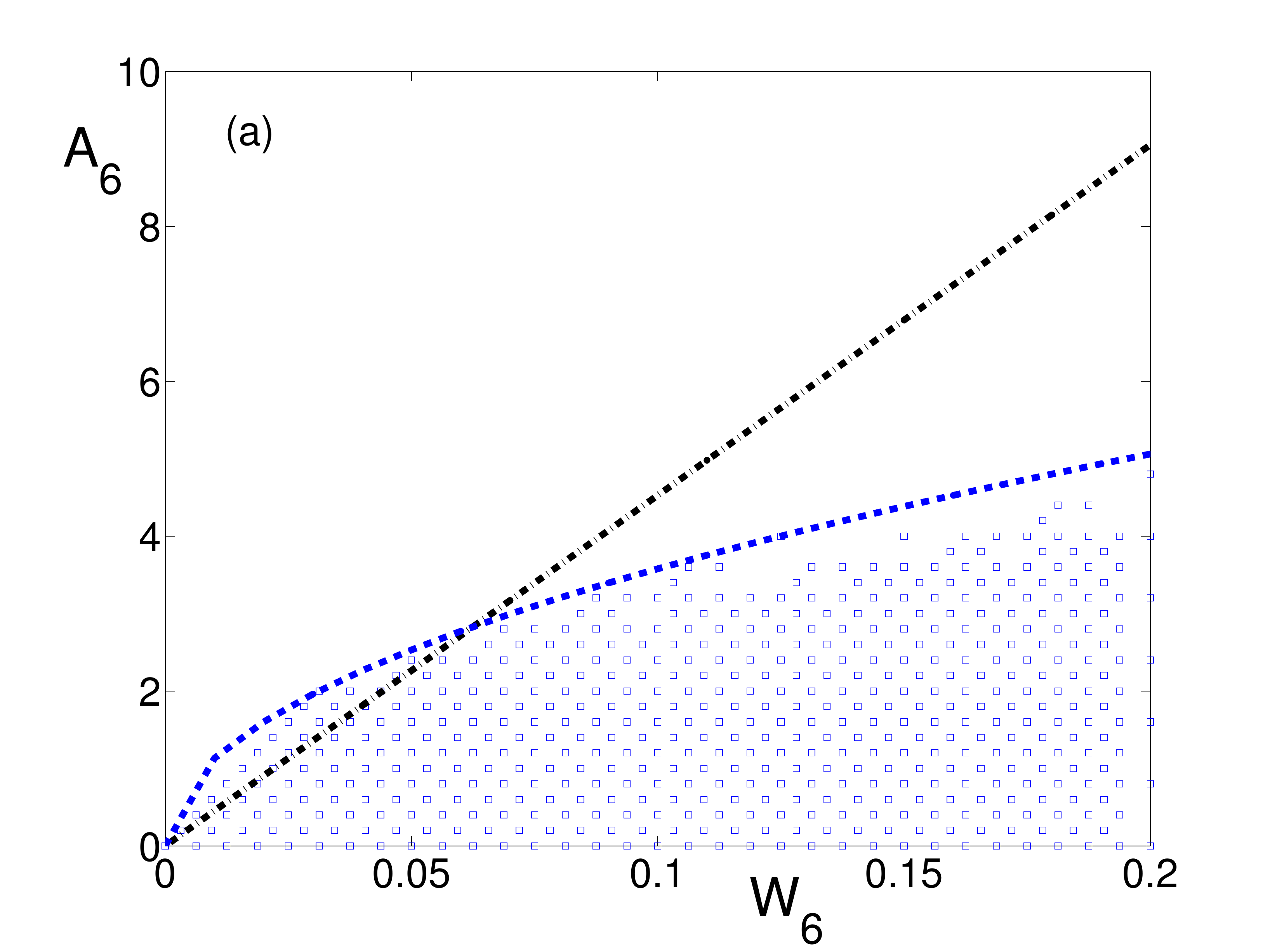} 
\par\end{centering}

\begin{centering}
\includegraphics[width=1\columnwidth]{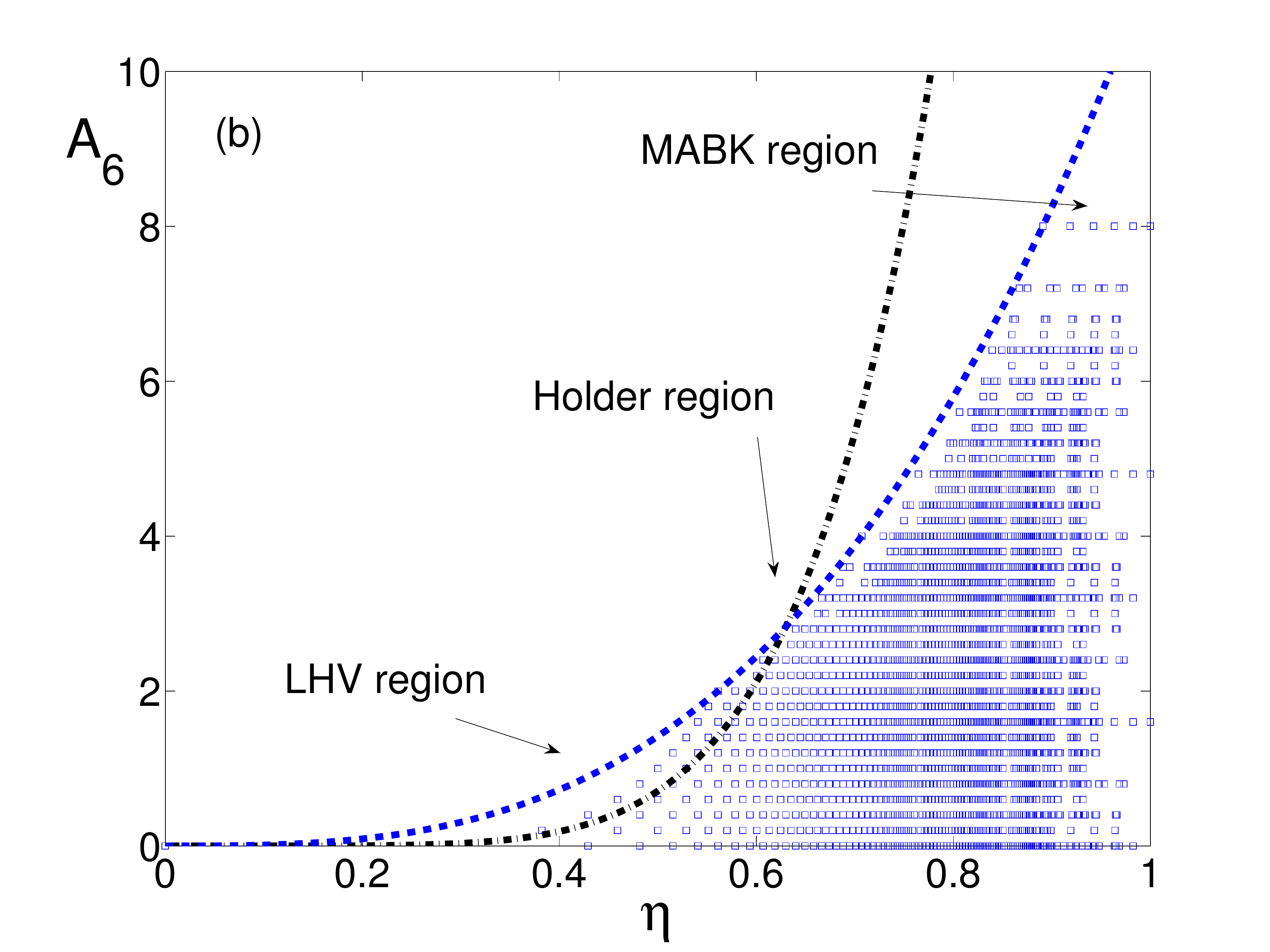}
\par\end{centering}

\caption{The LHV and quantum predictions for $Ar_{6}$ versus efficiency. The
curves are as defined in Figures 1, 2 and 3. The top figure gives
the predictions for $Ar_{6}$ for a given value of $W_{N}$. The lower
figure gives the predictions for the special symmetric case, where
$\eta_{k}=\eta$, $\forall k$. The blue curves are the Holder and
MABK inequality upper bounds to the LHV predictions. The black line
is the quantum prediction for a GHZ state. The Holder region is clearly
apparent and in this region, we see that the Holder curve is a close
fit to actual LHV predictions, indicated by the blue squares. The
quantum prediction violates the LHV one, for $\eta$ approaching the
limit of $0.5$, in this region.}
\end{figure}

More interesting behaviour is noticed for higher $N$. We identify
three regions.

\textbf{(1) MABK region}: The figures show the region defined by $W_{N}>0.5$
(which corresponds to $\eta>0.5^{1/N}$ in the symmetric case where
each $\eta_{k}=\eta$) and for which the LHV bound is that given by
the MABK Bell inequalities. In this parameter range of $W_{N}$, which
we call the {}``MABK region'', the Holder inequality bound is irrelevant.
This region cannot be reached unless the efficiency \emph{at each}
\emph{site }exceeds 50\%: ie. $\eta_{k}>0.5$. 

For $W_{N}<0.5$, we classify two regions. 

\textbf{(2) LHV region:} This region is defined by $0<W_{N}<0.5^{N}$,
which requires in the symmetric case, where all $\eta_{k}$ are equal,
that the efficiency at each site is below $50\%$. This may be thought
of as a {}``no-violation'' or {}``LHV region'' in that case, because
of the simple result, that LHV theories cannot be violated using two-setting
inequalities, if $\eta_{k}<0.5$ for each $k$. We outline an intuitive
proof.

\textbf{\emph{Proof}}: Suppose $N=2$, and that measurements at each
site are made by observers Alice and Bob, respectively. Suppose also
that losses are 50\% at each of Alice's and Bob's channel. It is then
possible that an {}``Eve'' has tapped into Alice's channel using
a 50:50 beam splitter, and has created a second channel symmetric
to Alice's. Eve can make measurements on this second channel, simultaneously
to Alice's measurements. Alice can choose to measure either $A_{1}$
or $B_{1}$, and Eve can choose to measure either $A_{1}^{E}$ or
$B_{1}^{E}$. In this case, by symmetry, we deduce that Eve's measurements
can have the same correlation with the measurements made by Bob as
Alice's measurements. A similar second Eve can exist at Bob's channel,
at site $k=2$. This second Eve can make measurements $A_{2}^{E}$,
$B_{2}^{E}$. The potential existence of the two Eves necessarily
downgrades the correlations of Alice and Bob, measured by $P(A_{1},A_{2})$,
$P(B_{1},A_{2})$, $P(A_{1},B_{2})$ and $P(B_{1},B_{2})$, so that
the probabilities cannot give a violation of the two-setting Bell
inequality. This follows, since $A_{1}$, $A_{2}$, $B_{1}^{E}$ and
$B_{2}^{E}$ can be measured simultaneously, and therefore there exists
a joint probability distribution for those outcomes. The set $P(A_{1},A_{2})$,
$P(B_{1}^{E},A_{2})$, $P(A_{1},B_{2}^{E})$ and $P(B_{1}^{E},B_{2}^{E})$
cannot therefore violate the Bell inequality. The symmetry of the
correlations ($P(A_{1},B_{2})=P(A_{1},B_{2}^{E})$, etc) then allows
us to deduce that there can be no violation of the Bell inequality
for the measurements of Alice and Bob (since there exists an underlying
probability distribution for these outcomes). The result is readily
extended to higher $N$. 

The proof depends on the existence of a symmetric beam splitter that
creates, from one channel, $2$ symmetric channels, to give 50\% loss
on the first channel. The proof also utilises that the Bell inequality
involves just two settings at each site, so that simultaneous measurements
performed on two channels at each site can completely specify a joint
probability distribution for the Bell inequality. An extension of
the argument, assuming existence of a device that creates $m$ symmetric
channels from $1$ channel, would lead to the conclusion that an $m$-setting
Bell inequality cannot be violated where $\eta_{k}\leq1/m$. Thus
we deduce the requirement of $\eta_{k}>1/m$ for at least one $k$,
for violation of an $m$-setting Bell inequality$\,\Diamond$. 

The region $W<0.5^{N}$ is evident in the Figures as that corresponding
to a straight-line relationship between actual LHV predictions and
$W_{N}$. As expected, the quantum prediction is within the bound
set by the LHV predictions. 

\textbf{(3) Holder region: }The next region is the most interesting
to us. This region shows a different LHV curve, closely approximated
by the Holder analytic bound. We call this the {}``Holder'' region.
On examining the Figures, we find that, as $N$ increases above $2$,
the quantum prediction moves from the MABK region ($W_{N}>0.5$) to
intersect the LHV bound in the Holder region. This allows an analytic
expression for the threshold efficiency in order to violate the two-setting
Bell inequality:
\begin{equation}
\prod_{k=1}^{N}\eta_{k}>2^{(2-N)}\label{eq:newthres}
\end{equation}
 \textcolor{black}{which corresponds to $\eta=0.79$ }for $N=3$ and
$\eta=0.5$ in limit of larger $N$, in the symmetric case $\eta_{k}=\eta$. 

Our analysis thus establishes three new results. The main result is
that the Holder expression gives a close fit to the LHV predictions,
in this Holder region. This provides an analytical tool for understanding
the LHV bounds in the two-setting scenario with loss. Second, we note
that the quantum GHZ prediction intersects the Holder LHV bound, for
all $N\geq3$ even and odd, and moves {}``down'' toward the edge
of the {}``no-violation LHV'' region as $N\rightarrow\infty$. The
third new result is that violation of the two-setting Bell inequalities
can be obtained without the requirement that \emph{each} $\eta_{k}$
be greater than 50\% (provided $N>3$). This is evident from the efficiency
threshold (\ref{eq:newthres}). We see that if $N-1$ efficiencies
are $1$, we only need an efficiency $\eta_{1}>4/2^{N}$ at the remaining
site for a violation of the Bell inequality. This efficiency $\eta_{1}$
can be vanishingly small as $N\rightarrow\infty$.

\section{Discussion }

The predicted efficiency thresholds do not quite match those shown
to be possible by Cabello, Rodriguez and Villanueva \cite{cabello-1}
for the case of odd $N$, but come very close (for $N=3$, $0.79$
versus $0.75$, for $N=5,$ $0.61$ versus $0.58$, for $N=7$). The
difference is that CRV imposed an additional symmetric constraint
on the LHV model, that each individual $\eta_{k}$ is measured and
found precisely equal ($\eta_{k}=\eta$). This condition is practically
reasonable, but is not imposed here. We have conditioned only on the
value of $W_{N}$. Our case is informative, however, in revealing
low efficiency thresholds in the asymmetric case, without the assumption
of symmetric sites, and provides a rigorous way to test Bell nonlocality
loophole-free, for practical realisations involving asymmetric transmission
of entangled qubits.

The efficiency bounds deduced by Larsson and Semitecolos \cite{LS}
are even lower for a specified $N$, but are obtained using Clauser-Horne
inequalities and nonmaximally entangled states. While CH inequalities
are useful for loophole-free Bell tests \cite{CH,cs}, they rely on
rarer joint detection events and thus is usually a less efficient
use of the resource, particularly for larger $N$ \cite{parampNdemart}.
Understanding how to test loophole-free Bell nonlocality for the
MABK situation and for maximally entangled GHZ states is therefore
an important goal. 

On that note, it is interesting to conjecture the usefulness of the
Bell nonlocality that is realised in the two different regions, MABK
and Holder. For many quantum information tasks, it is the genuine
$N$-partite form of nonlocality that is the most useful \cite{ssgen}.
Genuine Bell nonlocality was considered by Svetlichny \cite{svet},
and requires that the Bell nonlocality be truly shared among all $N$
sites, so that, for example, the system is not describable by the
Bell nonlocality of a $k$-partite GHZ state, where $k<N$. The best
known criterion for genuine Bell nonlocality is a violation of the
Svetlichny inequality \cite{svet}. This inequality reduces to $Ar_{N}\leq2^{N-1}$
in our notation, and requires $W_{N}>\frac{1}{\sqrt{2}}$ ( $\eta>(\frac{1}{\sqrt{2}})^{1/N}$
for symmetric efficiencies) for violation, a violation that can only
be obtained in the MABK region.  We remark that any more general
criterion for genuine Bell nonlocality will require, at least, that
$\eta_{k}>0.5$ for each site. This remark is based on the result
that Bell nonlocality will always imply a type of nonlocality called
{}``steering'' \cite{steer,eprsteercav}. From this knowledge, one
may utilise results of Ref. \cite{eprsteer} to establish the requirement
of $\eta_{k}>0.5$ for each site. This requirement, however, does
not necessarily imply the MABK region, and we leave as an open question
whether genuine multipartite Bell nonlocality can be observed loophole-free
in the Holder region.

\section{Conclusion}

In summary, we have established that the threshold efficiency for
failure of local realism using GHZ states and the correlations of
the two-setting MABK inequalities is $\prod_{k=1}^{N}\eta_{k}<0.5$,
where $\eta_{k}$ is the efficiency at the $N$th site. This means
that the maximally entangled GHZ state can indeed violate the predictions
of LHV models, for symmetric efficiencies $\eta\rightarrow0.5$ as
$N\rightarrow\infty$. Furthermore, we have shown that for two-setting
inequalities, there is no requirement (for loophole-free Bell tests)
that the efficiency $\eta_{k}$ at \emph{each} site exceed 50\%, provided
$N>3$. 

The proposed experiment is very simple, and requires a measurement
of efficiency only by measurement of the correlation $W_{N}$ which
is readily evaluated from the spin results. While $\eta>0.5$ is
a challenge for current experiments involving photons, the approach
developed here may be extended to multi-setting Bell inequalities,
for which the fundamental efficiency constraint is lower than $0.5$.
The inequalities could be useful for detecting Bell nonlocality in
future heralded experiments involving material particles, where loss
is determined to be at a level somewhere between 0.5 and 1.
\begin{acknowledgments}
I acknowledge support from the ARC Discovery Project Grants scheme
and stimulating discussions with P. Drummond, Q. He and S. Kiesewetter. \end{acknowledgments}

\end{document}